\documentclass[aps,prd,preprintnumbers,showpacs,showkeys,nofootinbib,
superscriptaddress,fleqn,floatfix,tightenlines,10pt]{revtex4-2}
\usepackage{amsmath,amsfonts,amssymb,amscd,amsxtra,amsthm}
\usepackage{graphicx}  
\usepackage{epstopdf}
\usepackage{dcolumn}  
\usepackage{bm}          
\usepackage{slashed}
\usepackage{cancel}
\usepackage{float}
\usepackage{mathtools}
\usepackage{amsbsy}
\usepackage{amstext}

\usepackage[utf8]{inputenc} 
\usepackage{booktabs} 
\usepackage[normalem]{ulem} 
\usepackage[dvipsnames]{xcolor} 
\usepackage{tabularx}
\usepackage{enumitem}  
\usepackage{array} 
\usepackage{slashed}
\usepackage{tikz}
\usepackage{float}
\usepackage{multirow}
\usepackage{cancel}
\renewcommand\sout{\bgroup \color{red} \ULdepth=-.5ex \ULset}

\makeatletter

\begin{document}  
\title{Parametrization of transition energy-momentum tensor form factors} 
\author{June-Young Kim}
\email[E-mail: ]{Jun-Young.Kim@ruhr-uni-bochum.de}
\affiliation{Institut f\"ur Theoretische Physik II, Ruhr-Universit\"at
  Bochum, D-44780 Bochum, Germany}

\date{\today}
\begin{abstract}
While the Lorentz structures of the $N \to \Delta$ transition matrix element of the vector current is well-known, that of the symmetric energy-momentum tensor current is unknown. In this letter, we first reproduce the Lorentz structures of the $\frac{1}{2}^{\pm} \to \frac{1}{2}^{\pm}$, $\frac{1}{2}^{\mp} \to \frac{1}{2}^{\pm}$, and $\frac{1}{2}^{\mp} \to \frac{3}{2}^{\pm}$ transition matrix elements of the vector current. We also repeat the parametrizations of the symmetric energy-momentum tensor current for the $\frac{1}{2}^{\pm} \to \frac{1}{2}^{\pm}$ and $\frac{1}{2}^{\mp} \to \frac{1}{2}^{\pm}$ transitions. We then consider both $\frac{1}{2}^{\pm} \to \frac{3}{2}^{\pm}$ and $\frac{1}{2}^{\mp} \to \frac{3}{2}^{\pm}$ matrix elements of the symmetric energy-momentum tensor current (say, $N \to \Delta$ transition) for the first time. We found that there are five independent conserved and four independent non-conserved energy-momentum tensor form factors. 
\end{abstract}
\maketitle

\section{Introduction}
The hadronic matrix element of the local current is in general parametrized in terms of the form factors. The representative examples are the electromagnetic~(EM) and energy-momentum tensor~(EMT) form factors. These form factors include the fundamental information of the internal structure of the hadrons. From the EM form factor, one can access the information on the charge and magnetization distributions inside a hadron, and the EMT form factors provide not only how the mass and spin of partons are distributed inside a hadron but also how the mechanical forces act inside the hadron. While the EM form factor has been intensively investigated well over decades, the EMT form factors were merely considered as an academic subject~\cite{Kobzarev:1962wt, Pagels:1966zza}, since the graviton weakly interacts with matter. However, the modern understanding of the EM and EMT form factors as Mellin moments of the generalized parton distributions~(GPDs) provides an integrated picture of these form factors and allows us to extract the EMT form factor from the GPDs~\cite{Muller:1994ses, Ji:1996ek, Radyushkin:1996nd} (see also relevant reviews~\cite{Goeke:2001tz, Diehl:2003ny, Belitsky:2005qn}). Recently, there has been huge progress in the understanding of the nucleon EMT form factors on both the theoretical~\cite{Polyakov:2002yz, Polyakov:2018zvc, Lorce:2018egm} and experimental~\cite{Burkert:2018bqq, Burkert:2021ith} sides. For the higher-spin particle, such a study has been also carried out on the theoretical side~\cite{Cotogno:2019vjb, Cosyn:2019aio, Polyakov:2019lbq, Kim:2020lrs}. 

Extracting $N^{*} \to N$ transition GPDs is a recent target of the Jefferson Lab~(JLab), and the transition EMT form factor of the nucleon excited states can be determined from the exclusive electroproduction data~\cite{Proceedings:2020fyd}. For example, very recently, the hard exclusive $\pi^{-}$ production was considered to obtain the $p\Delta^{++}$ transition GPDs~\cite{Joo}. $N^{*}(1535)\to N$ transition EMT form factors were first considered in Ref.~\cite{Ozdem:2019pkg} and their complete expression was established in Refs.~\cite{Polyakov:2020rzq, Azizi:2020jog}. However, $\Delta(1232)\to N$ transition EMT form factors have not been studied at all. Moreover, the connection between the transition GPDs and transition EMT form factors is unknown. In this letter, we thus aim at providing the possible Lorentz structures of the transition matrix element of the EMT current.

\section{Symmetries}
The transition matrix element of the local current is written as
\begin{align}
\langle N^{*}, p', \sigma' | \hat{O}(0) |N, p, \sigma \rangle = \bar{u}^{\alpha}(p', \sigma') O_{\alpha} (P,\Delta) u(p, \sigma).
\end{align}
The matrix element depends on the spin polarizations $\sigma$ and $\sigma'$, the average momentum $P= (p+p')/2$ of the initial and final state, and the momentum transfer $\Delta= p'-p$. The squared of this momentum transfer is denoted by $t=\Delta^{2}$. The on-shell conditions of the final and initial four momenta are given by $p'^{2}=m^{2}_{f}$ and $p=m^{2}_{i}$, respectively, where $m_{f}$ and $m_{i}$ denote the masses of the final and initial states. The operator $\hat{O}$ indicates the vector $ \hat{O}^{\mu}_{V}=\bar{\psi}\gamma^{\mu} \psi$ or symmetric tensor $\hat{O}^{\mu \nu}_{T}= \bar{\psi}i \overleftrightarrow{D}^{ \{ \mu} \gamma^{\nu \} } \psi$ (say, quark part of the symmetric Belinfante-Rosenfeld EMT current) currents. Because of the Lorentz invariance, the Lorentz tensor $O(P,\Delta)$ can be expressed in terms of the metric tensor $g_{\mu \nu}$ and the four-rank Levi-Civita tensor $\epsilon^{\mu \nu \rho \sigma}$ and the four-vectors $P$ and $\Delta$. The dynamical information of the hadronic system is included in a Lorentz invariant function, i.e., form factor. The equation of motion of the Dirac spinor is given by
\begin{align}
\slashed{p} u(p,\sigma) = m u(p,\sigma).
\end{align}
The spin-3/2 spinor, i.e., Rarita–Schwinger spinor $u^{\alpha}$, satisfies the following subsidiary conditions:
\begin{align}
\gamma^{\alpha} u_{\alpha}(p,\sigma)=0, \ \ p^{\alpha} u_{\alpha}(p,\sigma)=0, \ \ \slashed{p} u_{\alpha}(p,\sigma)=m u_{\alpha}(p,\sigma).
\end{align}
These conditions reduce the number of the possible Lorentz structure of $O(P,\Delta)$.

In addition, the space-time discrete symmetries~(e.g., time-reversal, parity, and hermiticity) also reduce the number of the possible Lorentz structure and guarantee that the form factors are real-valued functions. Note that once these symmetries are relaxed additional form factors appear~\cite{Kobzarev:1962wt, Kobsarev:1970qm, Ng:1993vh, Polyakov:2020rzq}.
Under the parity and time-reversal transformation, the transition matrix element $(1/2 \to 3/2)$ should comply with the given relations for both the vector and symmetric tensor currents:
\begin{align}
O^{\mu}_{V\alpha}(P,\Delta)&=  -\gamma^{0} O^{\bar{\mu}}_{V\bar{\alpha}}(\bar{P},\bar{\Delta}) \gamma^{0} \eta_{f}\eta_{i}, \ \ \ \  O^{\mu}_{V\alpha}(P,\Delta)=  i\gamma^{1} \gamma^{3} O^{\bar{\mu}*}_{V\bar{\alpha}}(\bar{P},\bar{\Delta}) i\gamma^{1} \gamma^{3}, \cr
O^{\mu \nu}_{T\alpha}(P,\Delta)&=  -\gamma^{0} O^{\bar{\mu} \bar{\nu}}_{T\bar{\alpha}}(\bar{P},\bar{\Delta}) \gamma^{0} \eta_{f}\eta_{i},  \ \ \ \ 
O^{\mu  \nu}_{T\alpha}(P,\Delta)=  i\gamma^{1} \gamma^{3} O^{\bar{\mu} \bar{\nu}*}_{T\bar{\alpha}}(\bar{P},\bar{\Delta}) i\gamma^{1} \gamma^{3},
\label{eq:1}
\end{align}
where $\eta_{i} (\eta_{f})$ is the intrinsic parity of the initial~(final) state. Here we use the convention $v^{\bar{\mu}} = \bar{v}^{\mu} = (v^{0},-\bm{v})$. For the transition matrix element $(1/2 \to 1/2)$, the same relations are satisfied except for the additional negative sign under the parity transformation in Eq.~\eqref{eq:1}. Note that while hermiticity provides a constraint on the number of the form factors in the non-transition matrix element, it does not put any restriction on the form factors for the transition case~\cite{Ng:1993vh}. For example, the transition matrix elements of a vector current for a spin-1/2 particle are given by
\begin{align}
&\langle N^{*},p',\sigma' | \hat{O}^{\mu} | N, p,\sigma \rangle = \bar{u}(p',\sigma') (O^{\mu})_{fi}  u(p,\sigma), \cr
&\langle N, p,\sigma | \hat{O}^{\mu} |  N^{*},p',\sigma' \rangle = \bar{u}(p,\sigma) (O^{\mu})_{if}  u(p',\sigma').
\end{align}
The hermiticity then yields the following condition:
\begin{align}
\langle N^{*},p',\sigma' | \hat{O}^{\mu} | N, p,\sigma \rangle^{*} = \langle N, p,\sigma | \hat{O}^{\mu} | N^{*},p',\sigma' \rangle.
\end{align}
For the different initial and final states~(off-diagonal components), the hermiticity merely gives the relation between the $i\to f$ process form factor $(O^{\mu})^{\dagger}_{fi}$ and its inverse $f\to i$ process form factor $(O^{\mu})_{if}$. So, it does not say anything about the reduction of the number of form factors. However, in the case of the equal states~(diagonal components), the hermiticity gives constraints between $(O^{\mu})^{\dagger}_{ii}$ and $(O^{\mu})_{ii}$.
So, the hermiticity in the non-transition matrix element for the spin-1/2 particle gives the following constraints~\cite{Meissner:2009ww, Cotogno:2019vjb}: 
\begin{align}
O^{\mu}_{V}(P,\Delta)&=  \gamma^{0} O^{{\mu} \dagger}_{V}(P,-\Delta) \gamma^{0}, \ \ \ \  
O^{\mu \nu}_{T}(P,\Delta)=  \gamma^{0} O^{{\mu} {\nu} \dagger}_{T}(P,-\Delta) \gamma^{0}.
\label{eq:hermiticity}
\end{align}

\section{Vector current}
In this Section, we start with a brief review of the $\frac{1}{2}\to \frac{1}{2}$ vector transition. By using the discrete symmetries, the vector current can be parametrized as the three independent form factors in general. Taking into account the current conservation, the possible number of the form factors is reduced to two. Parametrizations of the vector current for $\frac{1}{2}^{\pm} \to \frac{1}{2}^{\pm}$ and $\frac{1}{2}^{\pm} \to \frac{1}{2}^{\mp}$ transitions are given by
\begin{align}
O^{\mu (\pm)}_{V}(P, \Delta)&= \left[G^{ (\pm)}_{1}(t)\left(\gamma^{\mu} - \frac{m_{f}\mp m_{i}}{\Delta^{2}} \Delta^{\mu} \right)  +G^{ (\pm)}_{2}(t)\left(P^{\mu} - \frac{m^{2}_{f} - m^{2}_{i}}{2\Delta^{2}} \Delta^{\mu} \right)\right]\left( \begin{array}{c} 1 \\ \gamma_{5} \end{array}\right), 
\end{align}
where the superscripts $(+)$ and $(-)$ denote the processes $\frac{1}{2}^{\pm} \to \frac{1}{2}^{\pm}$ and $\frac{1}{2}^{\pm} \to \frac{1}{2}^{\mp}$, respectively. Note that the parity-flip transition requires the additional $\gamma_{5}$. These vector currents are parametrized  as the two independent form factors $G^{(\pm)}_{j} (j=1,2)$, which is consistent with those in Refs.~\cite{Korner:1976hv, Kubis:2000aa}. In the equal mass limit, i.e., $m_{f}=m_{i}$, we are able to recover the typical EM form factors of the nucleon. Here, one should bear in mind that the given form factors are dimensional quantities and can be defined as  dimensionless quantities by multiplying a mass scale factor, such as $t$ and $(m_{f}+m_{i})/2$. The choice of this factor is a matter of convention and does not affect the classification of the Lorentz structures at all.

Another well-known process is $\frac{1}{2}\to \frac{3}{2}$ or $N \to \Delta$ transition. Under the constraints of the discrete symmetries, the vector current is parametrized as the four independent form factors. In addition, the current conservation excludes one of the form factors. So, we have only three independent form factors. The general expression of this vector current is given by
\begin{align}
O^{\mu (\pm)}_{V \alpha}(P, \Delta)&= \bigg{[}G^{(\pm)}_{1}(t)\left(\Delta^{\mu} \Delta_{\alpha} -g^{\mu}_{\alpha} \Delta^{2}\right) + G^{(\pm)}_{2}(t)\left(P^{\mu} \Delta_{\alpha}- \frac{m^{2}_{f}-m^{2}_{i}}{2} g^{\mu}_{\alpha} \right) \cr
&+G^{(\pm)}_{3}(t) \left(\gamma^{\mu} \Delta_{\alpha} - g^{\mu}_{\alpha} (m_{f}\mp m_{i})\right) \bigg{]} \left( \begin{array}{c} 1 \\ \gamma_{5} \end{array}\right), 
\end{align}
where the superscripts $(+)$ and $(-)$ denote the processes $\frac{1}{2}^{\pm} \to \frac{3}{2}^{\mp}$ and $\frac{1}{2}^{\pm} \to \frac{3}{2}^{\pm}$, respectively. These matrix elements of the vector current are parametrized as the three different form factors $G^{ (\pm)}_{j} (j=1\ldots3)$, which is consistent with the results in Refs.~\cite{Jones:1972ky, Korner:1976hv}.

\section{Symmetric tensor current}
We are now in a position to discuss the symmetric EMT current. While the total EMT current is conserved, the separate quark $\hat{T}^{\mu \nu}_{q}$ and gluon $\hat{T}^{\mu \nu}_{g}$ EMT operators are not conserved unlike the vector current:
\begin{align}
\partial^{\mu} \hat{T}^{\mu \nu} =0, \quad \hat{T}^{\mu \nu}= \sum_{a=q,g}  \hat{T}^{\mu \nu}_{a}.
\end{align}
So, additional form factors emerge in the decompositions of their matrix elements. The conserved part is parametrized as the conserved EMT form factors and the non-conserved terms that break this conservation law are added in the separate quark and gluon level.

For the non-transition case, the EMT form factors have been intensively investigated in the series of papers~\cite{Polyakov:2002yz, Polyakov:2018zvc, Lorce:2018egm}. By using the discrete symmetries, the six independent EMT form factors are found. Moreover, having taken into account the current conservation and hermiticity~\eqref{eq:hermiticity}, two non-conserved form factors are excluded. In conclusion, the EMT current is parametrized as the four independent form factors. However, in the unequal mass or transition case (say, lowest-lying hyperon transition $\Lambda \to \Sigma$), the constraint on the number of the EMT form factors is relaxed, so that two additional non-conserved form factors appear, compared to the nucleon EMT form factors. In general, the symmetric tensor form factors are derived as
\begin{align}
O^{\mu \nu (\pm)}_{T} &= \bigg{[}F^{(\pm)}_{1}(t) \left( P^{\mu}P^{\nu}  + \frac{(m^{2}_{f}-m^{2}_{i})^{2}}{4\Delta^{2}} g^{\mu \nu} - \frac{m^{2}_{f}-m^{2}_{i}}{2\Delta^{2}} P^{ \{ \mu} \Delta^{\mu \}}  \right) \cr
&+ F^{(\pm)}_{2}(t) \left( \gamma^{ \{ \mu } P^{\nu \} } + \frac{(m^{2}_{f}-m^{2}_{i})(m_{f} \mp m_{i})}{\Delta^{2}} g^{\mu \nu}   - \frac{m_{f}\mp m_{i}}{\Delta^{2}} P^{ \{ \mu } \Delta^{\nu \} }  - \frac{(m^{2}_{f}-m^{2}_{i})}{2\Delta^{2}}  \gamma^{ \{ \mu } \Delta^{\nu \} } \right) \cr
&+F^{(\pm)}_{3}(t) \left( \Delta^{\mu}\Delta^{\nu} - \Delta^{2} g^{\mu \nu}\right)+\left( \bar{C}^{(\pm)}_{1}(t) g^{\mu \nu} +\bar{C}^{(\pm)}_{2}(t) \Delta^{ \{ \mu} P^{\nu \} } +\bar{C}^{(\pm)}_{3}(t) \gamma^{ \{ \mu} \Delta^{\nu \} }  \right) \bigg{]} \left( \begin{array}{c} 1 \\ \gamma_{5} \end{array}\right),
\end{align}
where the superscripts $(+)$ and $(-)$ denote the processes $\frac{1}{2}^{\pm} \to \frac{1}{2}^{\pm}$ and $\frac{1}{2}^{\pm} \to \frac{1}{2}^{\mp}$, respectively, and the symmetrization operator is defined as $X^{\{\mu  }Y^{ \nu \} } = X^{\mu} Y^{\nu}+ X^{\nu} Y^{\mu}$. There are three independent conserved EMT form factors $F^{(\pm)}_{j} (j=1 \ldots 3)$, and three independent non-conserved form factors $\bar{C}^{(\pm)}_{j} (j=1 \ldots 3)$. 
In the equal mass limit $m_{f}=m_{i}$, we have
\begin{align}
O^{\mu \nu (+)}_{T} &=F^{(\pm)}_{1}(t)  P^{\mu}P^{\nu}   + F^{(\pm)}_{2}(t)  \gamma^{ \{ \mu } P^{\nu \} }   +F^{(\pm)}_{3}(t) \left( \Delta^{\mu}\Delta^{\nu} - \Delta^{2} g^{\mu \nu}\right) \cr
&+\bar{C}^{(\pm)}_{1}(t) g^{\mu \nu} +\bar{C}^{(\pm)}_{2}(t) \Delta^{ \{ \mu} P^{\nu \} } +\bar{C}^{(\pm)}_{3}(t) \gamma^{ \{ \mu} \Delta^{\nu \} }.
\label{eq:eqmass}
\end{align}
For the non-transition process $\frac{1}{2}^{\pm} \to \frac{1}{2}^{\pm}$, the hermiticity~\eqref{eq:hermiticity} will additionally exclude the two non-conserved terms $\bar{C}^{(+)}_{2}(t)$ and $\bar{C}^{(+)}_{3}(t)$ in Eq.~\eqref{eq:eqmass}. Renaming the EMT form factors with help of the Gorden identity, one is able to identify them as the three conserved $F^{(+)}_{j} (j=1 \ldots 3)$ and one non-conserved $\bar{C}^{(+)}_{1}$ form factors of the nucleon. However, since the hermiticity does not give any restriction on the EMT form factors in the parity-flip transition, we found that there are three independent conserved EMT form factors $F^{(-)}_{j} (j=1 \ldots 3)$ and three independent non-conserved form factors $\bar{C}^{(-)}_{j} (j=1 \ldots 3)$, which coincides with the results from Ref.~\cite{Polyakov:2020rzq}.

In the case of the $\frac{1}{2} \to \frac{3}{2}$ transition, the promising project of extracting the transition GPDs from the CLAS data is ongoing~\cite{Proceedings:2020fyd, Joo}. Since the transition EMT form factors can be considered as a second Mellin moment of the transition GPDs in principle, the understanding of the transition EMT is necessary. Here we provide the general parametrization of the $1/2 \to 3/2$ transition matrix element of the EMT current for the first time. These EMT form factors are parametrized as the nine form factors with help of the discrete symmetries:
\begin{align}
O^{\mu \nu (\pm)}_{T\alpha} &= \bigg{[}F^{(\pm)}_{1}(t) \left( g^{ \{ \mu }_{\alpha} P^{\nu \} } + \frac{m^{2}_{f} - m^{2}_{i}}{\Delta^{2}} g^{\mu \nu} \Delta_{\alpha} - \frac{m^{2}_{f}-m^{2}_{i}}{2\Delta^{2}} g^{ \{ \mu }_{\alpha} \Delta^{\nu \} } - \frac{1}{\Delta^{2}}  P^{ \{ \mu } \Delta^{\nu \} } \Delta_{\alpha} \right) \cr
&+F^{(\pm)}_{2}(t) \left( P^{\mu}P^{\nu} \Delta_{\alpha} + \frac{(m^{2}_{f}-m^{2}_{i})^{2}}{4\Delta^{2}} g^{\mu \nu} \Delta_{\alpha} - \frac{(m^{2}_{f}-m^{2}_{i})}{2\Delta^{2}} P^{ \{ \mu} \Delta^{\nu \} } \Delta_{\alpha}  \right)  \cr
&+F^{(\pm)}_{3}(t) \left( \Delta^{\mu}\Delta^{\nu} - \Delta^{2} g^{\mu \nu}\right) \Delta_{\alpha}  \cr
&+F^{(\pm)}_{4}(t) \left( \gamma^{ \{ \mu } g^{\nu \} }_{\alpha} +\frac{2(m_{f}\mp m_{i})}{\Delta^{2}}g^{\mu \nu} \Delta_{\alpha}  - \frac{m_{f}\mp m_{i}}{\Delta^{2}} g^{ \{ \mu }_{\alpha} \Delta^{\nu \} } - \frac{1}{\Delta^{2}} \gamma^{ \{ \mu } \Delta^{\nu \} } \Delta_{\alpha} \right)  \cr
&+F^{(\pm)}_{5}(t)\left( \gamma^{\{ \mu} P^{\nu \} } \Delta_{\alpha} + \frac{(m^{2}_{f}-m^{2}_{i})(m_{f} \mp m_{i})}{\Delta^{2}}g^{\mu \nu} \Delta_{\alpha} - \frac{m_{f}\mp m_{i}}{\Delta^{2}} \Delta^{\{\mu } P^{\nu \} } \Delta_{\alpha} - \frac{(m^{2}_{f}-m^{2}_{i})}{2 \Delta^{2}}  \gamma^{\{ \mu } \Delta^{\nu \}}  \Delta_{\alpha}  \right)   \cr
&+\left(\bar{C}^{(\pm)}_{1}(t) g^{\mu \nu} \Delta_{\alpha} +\bar{C}^{(\pm)}_{2}(t) \Delta^{ \{ \mu} P^{\nu \} } \Delta_{\alpha}+\bar{C}^{(\pm)}_{3}(t) \gamma^{ \{ \mu} \Delta^{\nu \} } \Delta_{\alpha} + \bar{C}^{(\pm)}_{4}(t) g^{ \{ \mu}_{\alpha}  \Delta^{\nu \} } \right)  \bigg{]} \left( \begin{array}{c} 1 \\ \gamma_{5} \end{array}\right).
\end{align}
where the superscripts $(+)$ and $(-)$ denote the processes $\frac{1}{2}^{\pm} \to \frac{3}{2}^{\mp}$ and $\frac{1}{2}^{\pm} \to \frac{3}{2}^{\pm}$, respectively. We found that there are five independent conserved EMT form factors $F^{(\pm)}_{j} (j=1 \ldots 5)$ and four independent non-conserved form factors $\bar{C}^{(\pm)}_{j} (j=1 \ldots 4)$.

\section{Conculsion and outlook}

In this letter, we aim at providing the transition matrix element of the energy-momentum tensor current. We first examine the parity, time-reversal, and hermiticity to exclude the unphysical form factors.  In order to test the given symmetries, we considered the electromagnetic transition form factors and were able to reproduce them for both the parity-invariant  and -flip transitions. We then provided the transition energy-momentum tensor form factors $\frac{1}{2}^{\pm} \to \frac{1}{2}^{\pm}$ and $\frac{1}{2}^{\pm} \to \frac{1}{2}^{\mp}$ which coincide with the results in Ref.~\cite{Polyakov:2020rzq}. Finally, we presented the $\frac{1}{2} \to \frac{3}{2}$ transition energy-momentum tensor form factor for both the parity-invariant and -flip cases. We found that there are the five conserved $F^{(\pm)}_{j} (j=1 \ldots 5)$ and four non-conserved $\bar{C}^{(\pm)}_{j} (j=1 \ldots 4)$ energy-momentum tensor form factors. 

It is pretty straightforward to connect the transition energy-momentum tensor form factors to the transition GPDs as Mellin moments, which will provide a deeper understanding of the transition GPDs.

\begin{acknowledgments}
J.-Y. Kim is supported by the Deutscher Akademischer Austauschdienst~(DAAD) doctoral scholarship. J.-Y. Kim wants to express gratitude to  Cédric Lorcé, Bao-Dong Sun and Hyun-Chul Kim for the valuable comments.
\end{acknowledgments}

\end{document}